\documentstyle[aps,epsf]{revtex}  

\begin{document}        

\baselineskip 14pt
\title{Physics with the Main Injector}
\author{Hugh E. Montgomery}
\address{Fermi National Accelerator Laboratory, P.O. Box 500, Batavia, Illinois 60510, U.S.A.}
\maketitle              

\begin{abstract}        

 The Main Injector is a new rapid cycling accelerator at Fermilab
 which is a source of protons to be used in antiproton production to
 enhance the luminosity of the Tevatron Collider and to provide
 extracted beams for use in a range of fixed target experiments. We
 discuss the current status of the accelerator and the physics which
 it enables. The physics ranges broadly over the standard model and
 beyond, from the search for neutrino mass to collider physics at the
 highest energy available today.

\
\end{abstract}   	

\section{Introduction}               

  Since its construction the protons and anti-protons used in the
  Tevatron Collider have been produced and injected into the Tevatron,
  by the Main Ring which was the original 400 GeV proton synchrotron
  constructed during the 1970's. The Main Injector is a new 120-150
  GeV rapid cycling synchrotron in a tunnel separate from that of the
  Tevatron. The Main Injector improves significantly the anti-proton
  production capability of the complex. The new 8 GeV permanent magnet
  storage ring, the Recycler, in the Main Injector tunnel, permits
  reuse of the anti-protons that remain in the Tevatron at the end of
  a store.  Since most of the luminosity degradation results from beam
  dilution rather than anti-proton loss, this effectively doubles the
  available anti-protons for collisions.

  The greatly increased luminosity of the Tevatron Collider opens up
  windows on new physics. The production rate for heavy objects such
  as the top quark will be greatly enhanced. Perhaps most strikingly
  the increased luminosity may eventually provide sensitivity to the
  Higgs boson system which is thought to generate the mass of the
  observed particles, the $W$ boson, the $Z$ boson, and the quarks and
  leptons.

  The Main Injector is constructed with extracted beams that can be
  used for fixed target experiments. The operational possibilities are
  sufficiently flexible to interleave anti-proton production and fixed
  target physics with modest impact on either.  One of the primary
  uses of the high intensity extracted proton beam will be to produce
  a beam of neutrinos which will then be directed towards the Soudan
  mine in Minnesota where a distant detector can make measurements on
  the evolution of the neutrino species in the beam.

  The couplings of the three generations of quarks are described using
  the Cabibbo-Kobayashi-Maskawa flavor mixing matrix. A complete
  understanding this matrix does not yet exist. For example, the
  violation of Charge-Parity symmetry has been observed in the decay
  of neutral long-lived $K^{0}_{L}$ mesons but nowhere else. Kaon
  beams from the Main Injector promise key measurements in that
  system. The analogous investigation of the $b$-quark system is only
  beginning. The collider experiments CDF and D\O\ or/and a new
  dedicated experiment, BTeV, have ready access to large numbers of B
  hadrons, thus permitting extensive measurements beyond those being made
  at electron-positron B factories.

  The varieties of physics in these different sectors are all at or
  beyond the edge of our theoretical understanding and offer new
  insight on our world. In the rest of this paper we explore the
  machine and the enabled physics in a little more detail.

\section{The Machine}

  The Main Injector is a rapid cycling proton synchrotron with a
  circumference half that of the Tevatron. Protons are injected from
  the Booster machine to the Main Injector at 8 GeV and are
  accelerated to 120 GeV for either extraction to the anti-proton
  production target or to an external physics target. For injection of
  either protons or anti-protons into the Tevatron, the Main Injector
  ramps to 150 GeV.

  The machine performance parameters of interest for the particle
  physicist are given in Table~\ref{table-miparam}. In mixed modes of
  operation, the Main Injector can deliver $2.5$~$10^{13}$ protons to
  the experimental target and $5.0$~$10^{12}$ protons to the
  anti-proton production target every 2.87 secs. This causes a 15-20\%
  reduction of anti-proton production as a result of the increased
  cycle time. There are also potential improvements which could
  ultimately yield $5-10$~$10^{13}$ protons per cycle.

 \begin{table}
 \caption{Main Injector Performance Characteristics.}
 \begin{tabular}{l|ccc} 
              &$\overline{p}$ Production    &Fast Spill     &Slow Spill\\  
 \tableline 
 Energy(GeV)          &120            &120            &120\\
 Protons per Cycle    & $5.0$~$10^{12}$ & $3.0$~$10^{13}$  & $3.0$~$10^{13}$  \\ 
 Flat Top(secs)& 0.01 &0.01 &1.00\\
 Cycle Time(secs)& 1.47 & 1.87  & 2.87 \ 
 \label{table-miparam}
 \end{tabular}
 \end{table}

 The Collider luminosity is controlled by the total number of
 anti-protons available to accelerate and store. The anti-protons are
 produced by the 120 GeV protons incident on a nickel target. The
 produced anti-protons are focussed with a Lithium lens and collected
 in a debuncher ring at 8 GeV. From the debuncher they are transferred
 to the accumulator ring, also at 8 GeV, where they are cooled, thus
 producing stored anti-proton bunches. These are transferred to the 8
 GeV Recycler ring before acceleration in the Main Injector, and
 eventually the Tevatron, to the full energy.

 During a store, nuclear collisions cause some attrition while the
 beam-beam effects lead to an increase in the emittance of the
 beams. This effect dominates the reduction in luminosity over a
 period of hours. The effect is most strong on the anti-protons since
 the proton bunches are, in general, more intense than those of
 anti-protons. At the end of a store, which is usually defined by the
 luminosity having dropped to a few tenths of its initial value, the
 number of anti-protons has only reduced by a factor of two or
 less. These survivors can be decelerated through the Tevatron and
 Main Injector and captured in the Recycler. This is an 8 GeV
 permanent magnet machine equipped with stochastic cooling. It permits
 the recovery of emittances suitable for reinjection into the Tevatron
 and thus results in an effective factor of two enhancement to the
 number of available anti-protons under normal operation. The Tevatron
 Collider will operate with the luminosities indicated in
 Table~\ref{table-tevchar}

 \begin{table}
 \caption{Tevatron Collider Operating Characteristics.}
 \begin{tabular}{l|cccc} 
 &Bunch Spacing(nsec) & Inst. Luminosity($\rm{10^{31} cm^{2} sec^{-1}}$) &Interactions per crossing&Luminous Region(cm)\\  
 \tableline 
 Run Ib (1994-6) & 3500 & 1.6\tablenote{This was typical, the absolute record exceeded 2.5 $\rm{10^{31} cm^{2} sec^{-1}}$.} & 1-2 & 30 \\
 Run II (2000-3) & 396/132 & 10/20 & 1-2/1-2& 30/15 \\
 Run III(2004-7) & 132 & 50 & 5 & 15 \
 \label{table-tevchar}
 \end{tabular}
 \end{table}

 Further enhancements can be expected from the introduction
 of electron cooling in the Recycler and from the use of tune
 compensation in the Tevatron. A possible accumulation of integrated
 luminosity as a function of time is shown in
 Table~\ref{table-integL}. Before the full operation of the Large
 Hadron Collider at CERN, more than 10~$\rm{fb^{-1}}$ could be
 expected.

 \begin{table}
 \caption{Tevatron Collider Integrated Luminosity.}
 \begin{tabular}{l|cccccccc} 
 Year &2000&2001&2002&2003&2004&2005&2006&2007\\  
 \tableline 
 Peak Luminosity($10^{31} cm^{2} sec^{-1}$)& 5 & 10 &20 && 40 & 50 & 50 & 50 \\ 
 Integrated Luminosity($\rm{fb^{-1}}$)&0.5&1.0&2.0&&4.5&5.5&5.5&5.5\\  
 Accumulated Luminosity($\rm{fb^{-1}}$)&0.5&1.5&3.5&&8.0&13.5&19.0&24.0\  
 \label{table-integL}
 \end{tabular}
 \end{table}

 The Main Injector was being commissioned\cite{bd-opspage} at the time
 of the conference and that process has gone extremely well. In
 particular the machine is being operated with parameter values near
 those of the design and with relatively minimal use of correction
 elements. As for performance, it is already close to its intensity
 and cycle time goals. Meanwhile the Recycler is still in the final
 stages of installation. Beam has been passed through a fraction of
 its circumference.

\section{The Physics}

\subsection{Neutrino Mixing and mass}

 Over the years there have been speculations about whether or not
 neutrinos have identically zero mass. If not, one expects the weak
 eigenstates states to mix so that a pure beam of neutrinos of one
 species will evolve to contain an admixture of neutrinos of one or
 more other species. This phenomenon is called neutrino oscillation.
 The probability that a transition takes place is proportional to
 $\sin^{2}(1.27{\Delta{m}^2}{L}/E)$ where the difference in mass
 squared between the two neutrino states, $ {\Delta{m}^2}$, is
 measured in $(eV)^{2}$, the path length, L, in kilometres and the
 energy, E, in GeV. The strength of the oscillations is usually
 described by a factor $\sin^{2}{2\theta}$.
 
\begin{figure}[ht]	
\vskip -0.5 cm	
\centerline{\epsfxsize 2.8 truein \epsfbox{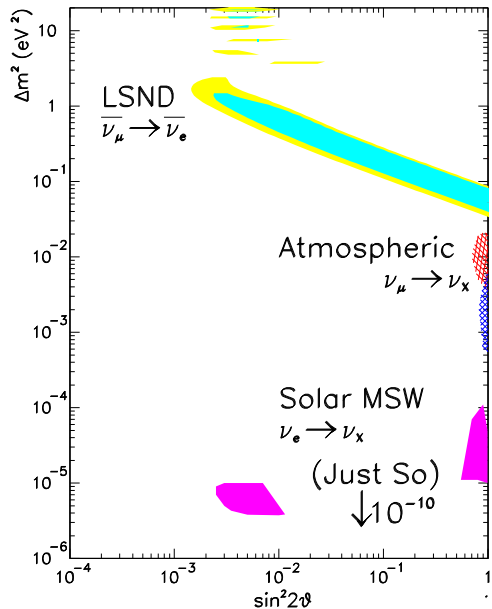}}   
\vskip -1.0 cm
\caption[]{
\label{fig-allhints}
\small Hints of neutrino oscillations from present measurements. Note that the most recent atmospheric  measurements, reported at this conference, suggest a region of ${\Delta{m}^2}$ somehwat higher than indicated in this figure from the summer of 1998. }
\end{figure}

 At the present time there are a number of observations\cite{feldman}
 from experiments which could be explained had the the neutrinos a
 finite mass. However the picture is quite complicated. The
 observation of a deficit of neutrinos from the sun suggests
 oscillations with very low $ {\Delta{m}^2}$. The experiments
 measuring the fluxes of atmospheric neutrinos, including the recent
 results from Kamiokande\cite{superkhere}, suggest
 ${\Delta{m}^2}\simeq 10^{-3} - 10^{-2}(eV)^{2}$ with maximal
 strength. The LSND experiment at Los Alamos has observed a hint of
 oscillations with $ {\Delta{m}^2}\simeq 10^{-2} - 10^{0}(eV)^{2}$
 with $\sin^{2}{2\theta}$ as low as $10^{-3}$. These observations are
 indicated in Fig.~\ref{fig-allhints}\cite{conrad_ichep}.

\begin{figure}[ht]	
\centerline{\epsfxsize 1.9 truein \epsfbox{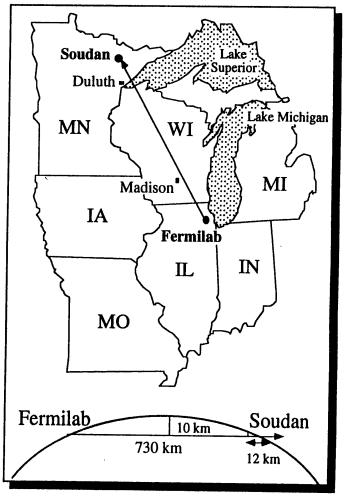}}   
\vskip 0 cm
\caption[]{
\label{fig-numimap}
\small Map indicating the trajectory of the neutrino beam from Fermilab to the Soudan mine in Minnesota. Inset at the bottom is a ``section'' of the earth showing the penetration of the neutrino trajectories through the earth's crust.}
\end{figure}

 The NuMI project, Neutrinos at the Main Injector, will construct a
 neutrino beam with energies peaking in the 10-25 GeV range. These
 neutrinos will be directed at two detectors, one on the Fermilab
 site, the other 740 km further north in the Soudan mine in Minnesota,
 see Fig.~\ref{fig-numimap}.

\begin{figure}[ht]	
\vskip -11.5 cm	
\centerline{\epsfxsize 3.3 truein \epsfbox{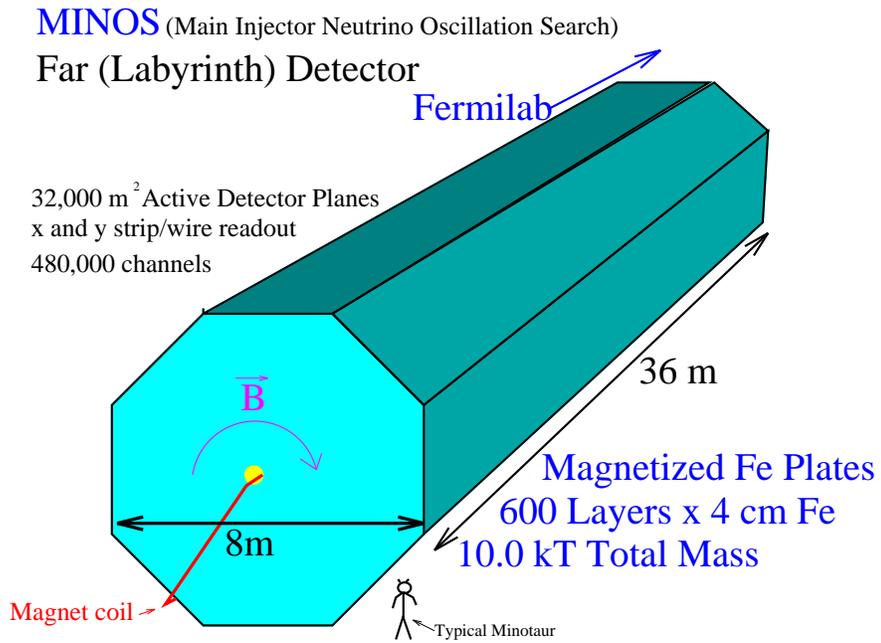 }}   
\vskip 9.5 cm
\caption[]{
\label{fig-minosfardet}
\small Layout of the MINOS ``Far Detector'' which will be situated in the Soudan mine.}
\end{figure}

 The MINOS experiment\cite{minos_expt}, Main Injector Neutrino
 Oscillation, will comprise two detectors one on the Fermilab site
 which monitors the neutrino beam interactions close to the source and
 one at the Soudan mine. The far detector sketched in
 Fig.~\ref{fig-minosfardet} will consist of iron toroid plates and
 solid scintillator sheets.  The goals of the experiment are
 unequivocally to observe the oscillations indicated by the
 SuperKamiokande and other experiments and to identify the mode(s) of
 oscillation. Neutral current events are distinguished from charged
 current events using measurements of the shapes of the neutrino
 induced showers. This technique gives some promise of positive
 identification of oscillations into $\nu_{\tau}$. This capability
 could be enhanced by a supplementary emulsion detector should the
 existing measurements continue to suggest that the $\nu_{\tau}$ mode
 is the relevant one. The sensitivity of the experiment is primarily
 at high values of $\sin^{2}{2\theta}$ and with $ {\Delta{m}^2}\ge
 10^{-3}(eV)^{2}$.

 The mini-BooNE experiment\cite{boone}, Booster Neutrino Experiment,
 is not a Main Injector experiment. Rather it uses protons from the 8
 GeV proton Booster machine to generate low energy neutrinos. Using an
 apparatus derived from that of LSND and situated about 1 km from the
 source, it aims definitively to cover that region of parameter space
 corresponding to the LSND observations. The systematic uncertainties
 would be significantly different from those of LSND.

\subsection{Physics of the Kaon System}

 The Cabbibo-Kobayashi-Maskawa(CKM) matrix has nine elements that can
 be described using four real parameters of which one is a phase
 angle. In turn, if the matrix is unitary, these parameters can be
 represented by a triangles. The lengths of the sides are controlled
 by the various transition amplitudes. The magnitude of
 Charge-Parity(CP) symmetry violation is controlled by the phase
 angle. This description of the quarks and their couplings may or may
 not hold in nature. It is one of the highest priorities of high
 energy physics to explore the CKM matrix in more detail and to
 determine whether or not the conjectures about its properties are
 true.

 At present the single indication of CP symmetry violation, which is
 what requires the complex matrix element, occurs in the $K^{0}_{L}$
 system\cite{blucher}. At present CP violation has not been observed
 in any other system containing strange quarks nor yet definitively
 observed in the $b$-quark sector\cite{cdf-sin2beta}.

 The conditions for existence of CP violation in any given system may
 be quite involved. In the $K^{0}_{L}$ system, for example, at the
 time of the conference it was still possible for the observed CP
 violation to be completely described by the mixing effects which are
 controlled by the parameter $\epsilon$. The search is for CP
 violation in the decay, `` direct CP violation'', which is controlled
 by the parameter $\epsilon$'. The new
 measurement\cite{blucher,ktev_new}, which appeared since the time of
 the conference, give $\cal{R}e(\epsilon$'/$\epsilon)$$\simeq (28 \pm
 4)\times10^{-4}$

\begin{figure}[ht]	
\centerline{\epsfxsize 5.5 truein \epsfbox{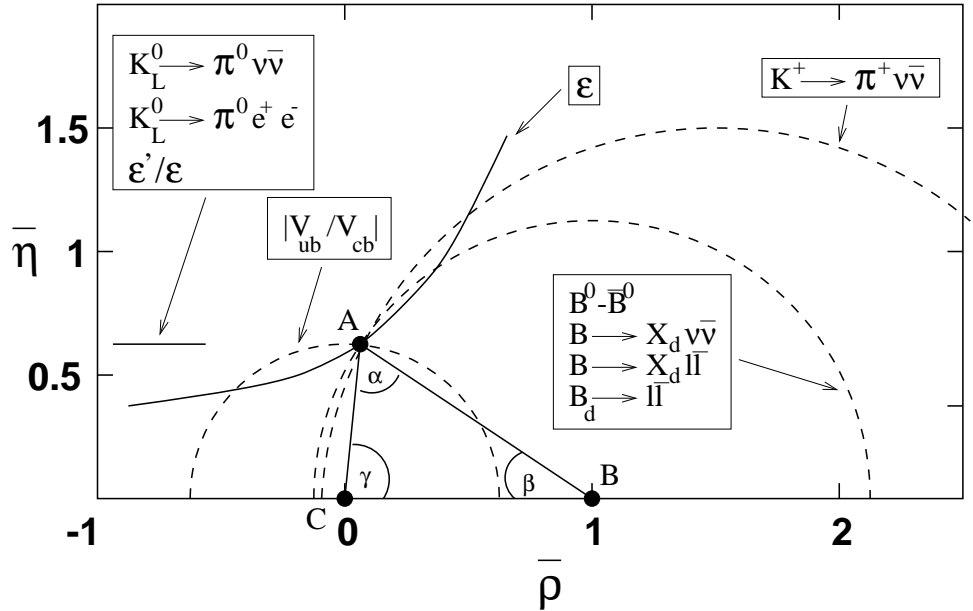}}   
\vskip -.2 cm
\caption[]{
\label{fig-burasfig}
\small The Unitarity Triangle associated with the Cabbibo-Kobayashi-Maskawa flavor mixing matrix and the measurements possible in the kaon system.}
\end{figure}

 In order to make progress, thoughts have turned to other
 possibilities\cite{cooper-kaons}. A version\cite{buras} of the
 unitarity triangle is shown in Fig.~\ref{fig-burasfig}. In principle
 it is possible to over-constrain the triangle and hence to test the
 theory using only measurements with kaons. As indicated, a
 measurement of the branching fraction for
 $K^{0}_{L}\rightarrow\pi^{0}\nu\overline{\nu}$ would directly
 constrain the height of the triangle while a similar measurement of
 the charged decay $K^{+}\rightarrow\pi^{+}\nu\overline{\nu}$
 determines the radius of an arc which should also pass through the
 apex of the triangle, if the theory is correct.

\begin{figure}[ht]	
\centerline{\epsfxsize 4.4 truein \epsfbox{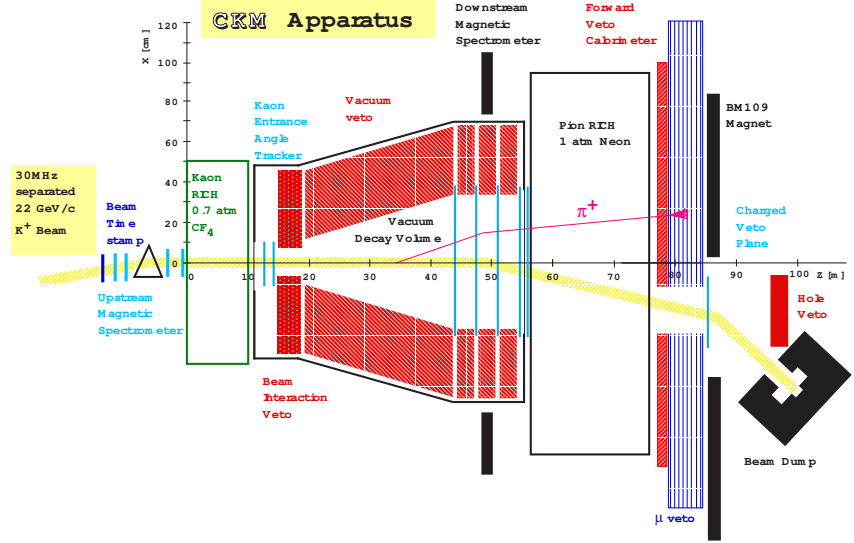}}   
\vskip 0 cm
\caption[]{
\label{fig-ckmexp}
\small Layout of the CKM Experiment.}
\end{figure}

 The charged kaon decay has been sought\cite{bnl-kaplus} at Brookhaven
 national Laboratory in the decays of stopped kaons with one event
 observed. Recently a proposal\cite{fnal-kaplus}, the ``CKM''
 experiment, has been made to use decays in flight. The apparatus is
 shown in Fig.~\ref{fig-ckmexp}; the beam would be a 22 GeV
 radiofrequency-separated beam of charged kaons at the Main
 Injector. It is interesting to look carefully at the aspect ratio of
 the experiment. It is very long and very narrow approximating an
 instrumented sewer pipe. The goal is to fully identify and measure
 the incident kaon and the outgoing charged pion, the only two
 measurable particles in the process.

 The equivalent neutral kaon experiment\cite{fnal-kami}, ``KaMI'',
 which would search for $K^{0}_{L}\rightarrow\pi^{0}\nu\overline{\nu}$
 could be derived from the KTeV experiment making the recent
 measurements of $\epsilon$'/$\epsilon$ at Fermilab. The key elements
 are the electromagnetic calorimeter and the photon vetos which are
 crucial to the background suppression. As with the charged kaon
 decay, this would be a very difficult measurement.

\begin{figure}[ht]	
\centerline{\epsfxsize 4.4 truein \epsfbox{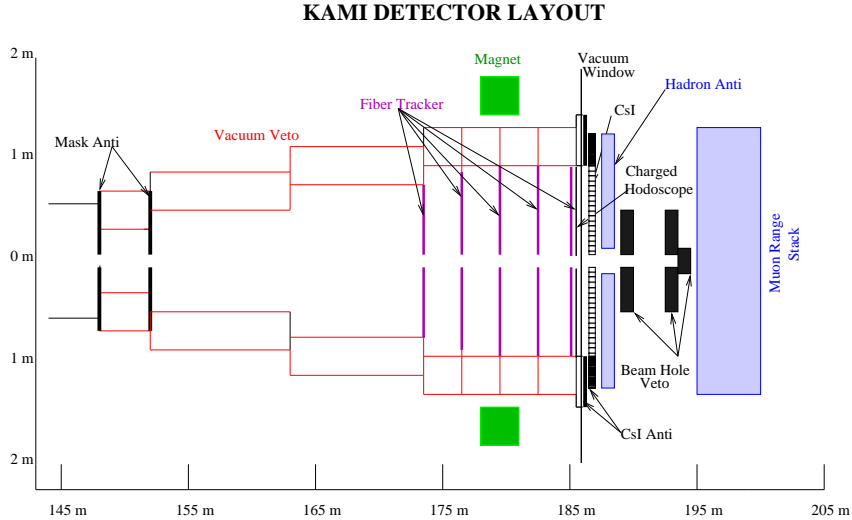}}   
\vskip 0 cm
\caption[]{
\label{fig-kamiexp}
\small Layout of the KAMI Experiment}
\end{figure}

\begin{figure}[ht]	
\centerline{\epsfxsize 2.5 truein \epsfbox{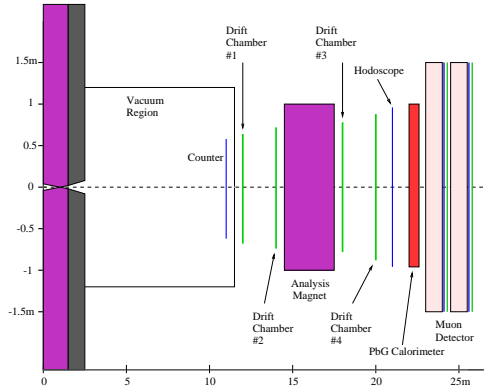}}   
\vskip 0 cm
\caption[]{
\label{fig-cptexp}
\small  Layout of the CPT Experiment.}
\end{figure}

 Completing the suite of kaon proposals is the ``CPT''
 Experiment\cite{fnal-cpt}. The goal is to measure CP violation in a
 number of modes especially in $K^{0}_{S}$ decays. It also gives the
 opportunity to measure the phase of the charged pion decays which in
 conjunction with the measurement of $\epsilon$'/$\epsilon$ gives a
 check on CPT symmetry with a sensitivity which corresponds to the
 Planck scale. It should be noted, see Fig.~\ref{fig-cptexp}, that
 since it is necessary to measure the interference terms between
 $K^{0}_{L}$ and $K^{0}_{S}$, the apparatus would be significantly
 shorter than either of the other two experiments. The $K^{0}$ beam is
 derived from the same RF-separated $K^{+}$ beam that is used for the
 ``CKM'' experiment.

\samepage

\subsection{Physics of the B System}

\begin{figure}[h]	
\vskip 0.2 cm
\centerline{\hspace{3cm}\epsfxsize 5.5 truein \epsfbox{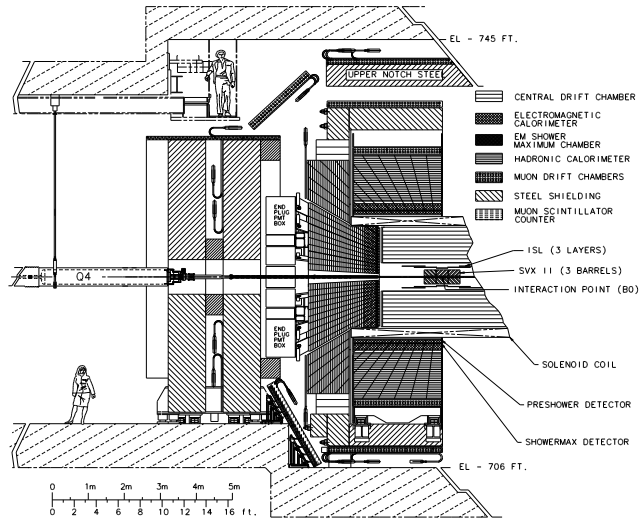}}   
\vskip 0.5 cm
\caption[]{
\label{fig-cdf}
\small The upgraded CDF detector. }
\end{figure}

\begin{figure}[h]	
\centerline{\hspace{1.5in}\epsfxsize 6.0 truein \epsfbox{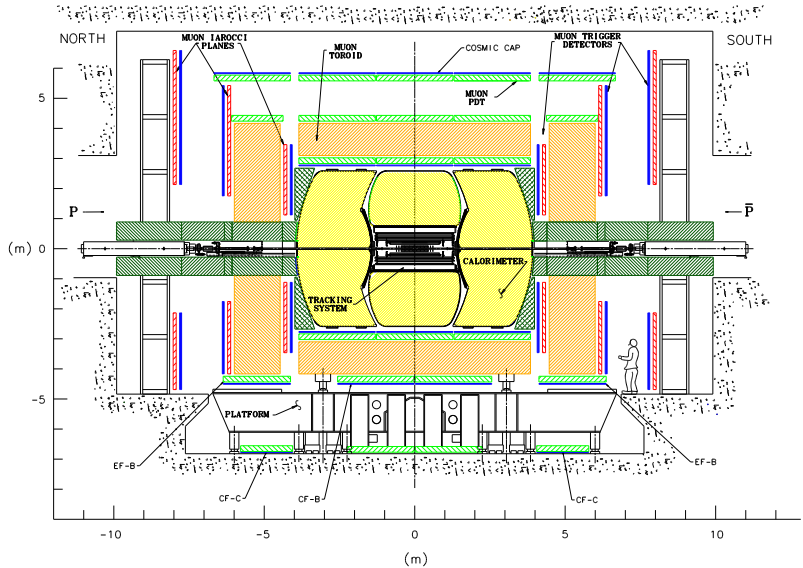}}   
\vskip -2 cm
\caption[]{
\label{fig-d0}
\small The upgraded D\O\ apparatus. }
\end{figure}

 In order to access large numbers of $b$ quarks at Fermilab, it is
 necessary to turn to the Tevatron Collider. The existing two
 detectors, CDF, see Fig.~\ref{fig-cdf}, and D\O\, see
 Fig.~\ref{fig-d0}, are being upgraded for operation in the Main
 Injector era. The upgrades\cite{cdf-upgrade,d0-upgrade} are extensive
 and are driven by the physics goals and by the changed operating
 characteristics of the Tevatron; the decreased spacing between bunch
 crossings has forced a rework of all the front-end eleectronic
 systems to introduce pipelines. A particular region of improvement is
 in the tracking in which each detector is being substantially
 modified.  D\O\ has installed a central solenoid and both experiments
 are constructing new outer trackers. The inner, silicon, systems will
 have upward of 600,000 channels each. These will provide high quality
 $b$-quark tagging and B meson reconstruction. In addition to
 enhancing the B-physics capabilities, $b$-quark tagging is recognised
 as an important tool in the exploration of and search for higher mass
 states. This was demonstrated in top-quark physics and is expected to
 be true for Higgs, SUSY or technicolor states.

 The physics which will be addressed by CDF and D\O\ is wide ranging.
 In later sections electroweak studies and physics beyond the standard
 model will be addressed but in this section we will concentrate on B
 physics. Recent results from CDF presage a bright future for the
 general purpose detectors at the Tevatron and for B physics at hadron
 colliders more generally. There is also an initiative to consider a
 dedicated detector\cite{btev-proposal}.

\begin{figure}[h]	
\centerline{\epsfxsize 4.4 truein \epsfbox{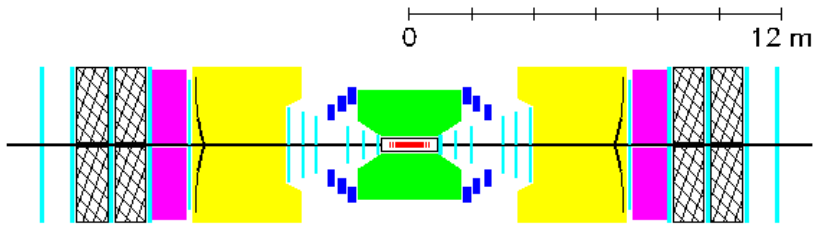}}   
\vskip 0 cm
\caption[]{
\label{fig-btev}
\small The BTeV apparatus. }
\end{figure}

 The BTeV proposal is motivated by the enormous production rate for
 $b$ quarks and the potential of a detector which accentuates the
 forward and backward directions in order to exploit the favorable
 mapping of rapidity to solid angle and to benefit from the relatively
 larger decay lengths for $b$ quarks of a given transverse
 momentum. The planned detector is shown in Fig.~\ref{fig-btev}. It
 consists of two spectrometer arms. Each arm is equipped with a
 silicon detector system inside the beam pipe and ring imaging
 Cherenkov counters. It is expected that the latter will be especially
 important.

  If the standard model is correct, the same triangle in Fig.~\ref{fig-burasfig} should describe the $b$-quark
  system. Measurements of the CP violating asymmetry in the decay $ B
  \rightarrow J/\psi K^{0}_{S}$ would determine $\sin 2\beta$. The
  recent measurement\cite{cdf-sin2beta} from CDF finds $\sin
  2\beta=0.79_{-0.44}^{+0.41}$ suggesting a positive value at about
  the 90\% C.L.  A feature of this measurement is the use of several
  different flavor tagging techniques. With approximately
  2$\rm{fb^{-1}}$ and the upgraded detectors, the uncertainty on $\sin
  2\beta$ will be reduced below 0.1 for each experiment. Similar
  uncertainties are projected for $\sin 2\alpha$ although  the interpretation for this case is considered to be more difficult. Measurement of the third angle, $\gamma$,  will be a challenge.

  During the last year, there have been a number of results from CDF
  on various aspects of higher mass B hadron states. The measurements
  of $B_s$ mixing\cite{cdf_b_s} are competitive with those from LEP
  and SLD. Extrapolating to Run II, we expect sensitivities in the
  range $x_s \ge 25$ from each experiment thus comfortably covering
  the expected range. The observation\cite{cdf_b_c} of the $B_c$ meson
  has further demonstrated that sophisticated studies of B physics are
  possible at a hadron collider. One should note that the final state
  used for this observation was semi-leptonic. Despite the incomplete
  reconstruction the backgrounds were manageable. Given the enormous
  rates, the Tevatron is arguably the best place to do B physics.
  This is especially true for states which are not decay products of
  the $\Upsilon_{4S}$ and are therefore difficult to access with an
  electron-positron B factory.

\subsection{Electroweak Physics}

  The hadron colliders have for some time contributed to the suite of
  measurements which provide stringent cross-checks of the electroweak
  model. The initial measurements from the CERN $Sp\overline{p}S$
  observation of the $W$ and $Z$ bosons paved the way. The current
  measurements\cite{wmass_measurements} from the Tevatron keep pace in
  precision with those from LEP and at each new level of precision we
  see ways to constrain the details of the production measurements
  from the data themselves. For example in the recent D\O\ measurement
  it was found that the constraint from the rapidity distribution of
  the bosons is only marginally weaker than that from the parton
  distribution function measurements from the worlds experiments. The
  current errors are 80-90 MeV per experiment. Already the latter are
  strongly influenced by measurements of the $W$ asymmetry at the
  Tevatron. The constraints from the data themselves will scale with
  statistics to higher integrated luminosity. This means that the
  current estimates of about 40 MeV per channel and per experiment are
  indeed possible.

\begin{figure}[]	
\centerline{\epsfxsize 2.6 truein \epsfbox{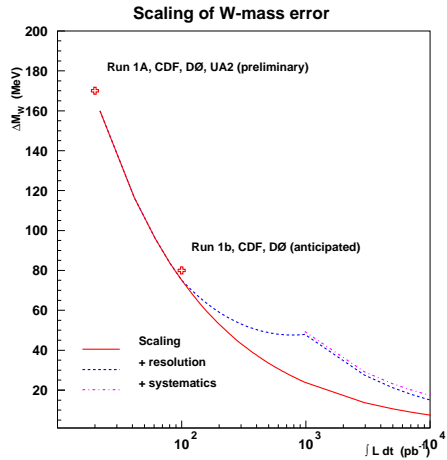}}   
\vskip 0 cm
\caption[]{
\label{fig-wmassevol}
\small Expected evolution of the precision of a measurement of the $W$-boson mass at the Tevatron Collider. }
\end{figure}

\begin{table}
 \caption{Uncertainties in a single experiment top mass determination.}
 \begin{tabular}{l|cc} 
 Uncertainty(GeV) & Run I & Run II\\  
 \tableline 
 Statistical& 5.6 & 1.3 \\
  \tableline
 Jet Energy Calibration & 4.0 & 0.4 \\
 Gluon ISR/FSR & 3.1 & 0.7 \\ 
 Detector Noise etc & 1.6 & 0.4 \\
 Fit Procedure & 1.3  & 0.3 \\
 \tableline
 All Systematic & 5.5 & 0.9 \\
  \tableline
 Total & 7.8  & 1.6 \
 \label{table-topmass}
 \end{tabular}
 \end{table}

  This kind of evolution as expressed in an earlier
  study\cite{tev2000} is illustrated in Fig.~\ref{fig-wmassevol}.
  That study considered particularly the effects of the underlying
  event on the technique which uses the transverse mass of the event
  (lepton plus neutrino) as the primary measure of the boson
  mass. This leads to a relative deterioration as the number of
  interactions per bunch crossing increases. What is also shown is the
  subsequent evolution as we decrease the bunch spacing in the machine
  to 132 nsecs. Fora given luminosity this decreases the number of
  interactions per crossing. We are also learning to take advantage of
  all the measures of the boson mass, the lepton transverse momentum,
  and the neutrino transverse momentum. It is gratifying that this
  projection made some four years ago holds good through the 100
  $\rm{pb}^{-1}$ of data from which the recent measurements are
  derived.


 A very powerful newcomer to precision measurements is the top
 mass. It enters into the electroweak parameters through its dominance
 of the quark loops. From the existing data, the 3\% uncertainty makes
 the top-quark mass the best known of all quark masses. The 5 GeV
 uncertainty in the top-quark mass is equivalent to about 30 MeV
 uncertainty in the $W$ mass in terms of its its sensitivity to the
 Higgs mass.. Currently the dominant error\cite{topmasserrors} comes
 from the calibration of the jets. Recently CDF observed the
 $Z\rightarrow b\overline{b}$ decay. Extrapolating to the luminosities
 and upgraded detectors with silicon track triggers to enhance the
 sensitivity to this channel, a very precise calibration of the jet
 energies becomes possible. This will be used in conjunction with the
 $W$-boson decays to jets which are present in the top signal data
 themselves. The resulting evolution of a single experiment in the
 lepton-plus-jets channel is illustrated in
 Table~\ref{table-topmass}. An uncertainty of less than 2 GeV appears
 to be possible.

\begin{figure}[h]	
\centerline{\epsfxsize 4.0 truein \epsfbox{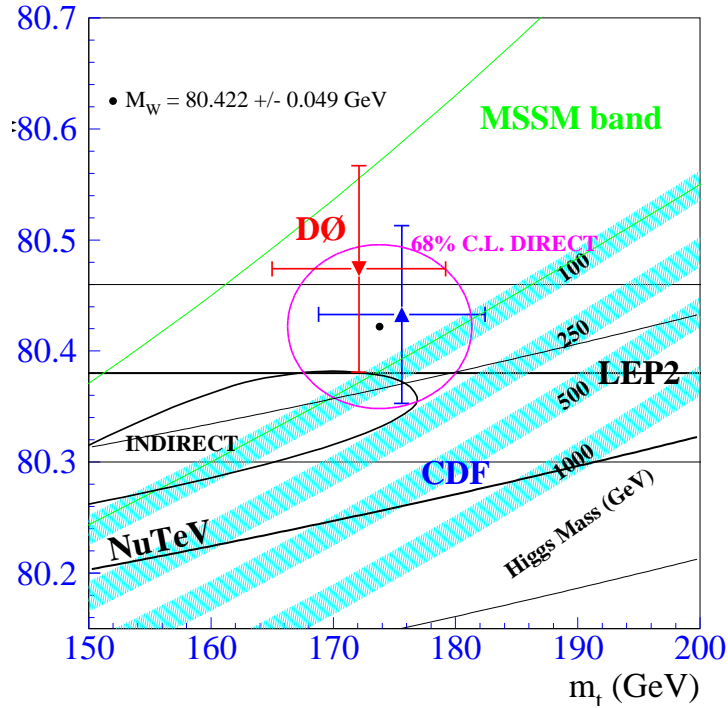}}   
\vskip 0 cm
\caption[]{
\label{fig-mwmt}
\small $M_{W}$ versus $M_{t}$. }
\end{figure}

  A number of other electroweak studies are possible including the
  comparison of the $W$ width as determined by direct and indirect
  methods and a study of the $Z$ asymmetry. The latter may serve
  either as a determination of $\sin^{2}\theta_{W}$ for the light
  quarks or another constraint on the parton distribution
  functions. Taken together, the masses of the $W$ boson and the top
  quark lead to constraints on the mass of the Higgs boson either in
  the standard model or in the supersymmetric variants. This is
  illustrated in Fig.~\ref{fig-mwmt} where, in the plane of the $W$
  mass and the top-quark mass, the various measurements including the
  latest from CDF and D\O\ at the Tevatron. We can anticipate that
  these indirect measurements will determine the Higgs mass to about
  50\% of its value.

\subsection{Physics beyond the Standard Model}

  There are as many searches for new physics as can be generated by
  the imagination of physicists. In the search for compositeness,
  structure, higher mass bosons, leptoquarks, the present limits are
  in the few-hundred-GeV range. With $2~\rm{fb}^{-1}$ these searches
  will be sensitive close to the 1 TeV range. Indirect searches as
  exemplified by the measurement of the Drell-Yan, lepton pair cross
  section and the dijet mass spectrum, are sensitive, through possible
  contact terms to the 5 TeV range.
  
\begin{figure}[]	
\centerline{\epsfxsize 3.3 truein \epsfbox{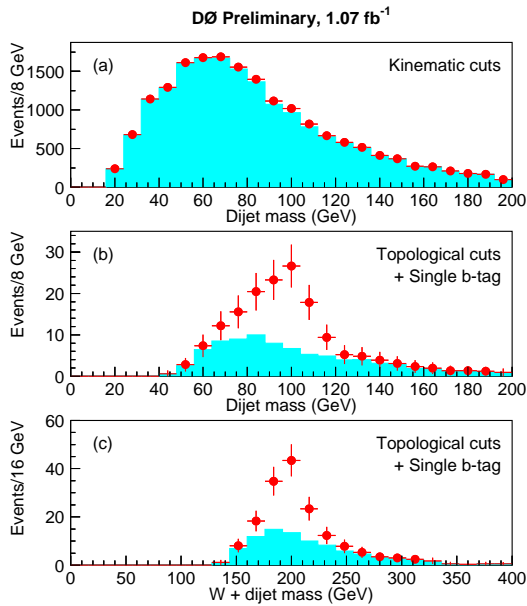}  }   
\vskip 0 cm
\caption[]{
\label{fig-technicolor}
\small Mass spectra expected for technicolor signals for the $\pi_{T}$ and the $\rho_{T}$ .}
\end{figure}

\subsubsection{Technicolor}

 A longstanding candidate to explain the origin of electroweak
 symmetry breaking is a new strong interaction, technicolor, analogous
 to QCD. Such a strong interaction would lead to massive electroweak
 bosons in a manner analogous to the way the masses of the pion and
 $\rho$ meson are generated by QCD. There are complications with the
 simplest forms of such a theory but
 variations\cite{technicolorsummary} of the original scheme continue
 to be explored.

\begin{table}[h]
 \caption{Mass ranges covered for a $5\sigma$ discovery in SUSY models.}
 \begin{tabular}{l|c|cc} 
 Model & SUSY Particle & Run I( 0.1 $fb^{-1}$ &  Run II(2.0 $fb^{-1}$\\
       &               & Mass Limit(GeV)      &    Mass Limit(GeV) \\
 \tableline 
 SUGRA &&\\
  \tableline
 & $\tilde{\chi}_{1}^{\pm}$ & 70\tablenote{95\% CL.}  & 210 \\
 & $\tilde{g}$ & 270\tablenote{95\% CL.} & 390 \\
 & $\tilde{t}_{1}(\rightarrow b\tilde{\chi}_{1}^{\pm})$& & 170 \\
  \tableline
 GMSB&&\\
  \tableline
 & $\tilde{\chi}_{1}^{\pm}$ &  & 265 \\
 & $\tilde{\tau}$ &  & 120 \
 \label{table-susy}
 \end{tabular}
 \end{table}

The production cross section for the some of the states can be quite
substantial\cite{womestilane}. Analysis of Monte Carlo simulations of
technicolor signals using a basic signature of a $W$, seen through its
leptonic decay, along with two jets gives a dijet mass spectrum as
shown in the upper plot of Fig.~\ref{fig-technicolor}. Some
topological requirements are then applied and a $b$-quark tag is
requested. This leads to the middle of the three plots. The dijet
spectrum clearly shows an excess which corresponds to the technipion
decaying to two $b$-quark jets. In the bottom plot, the mass of the
combined $W$ boson and the two-jet system shows a peak over background
corresponding to the technirho. Once again the importance of $b$-quark
tagging techniques is demonstrated.

\subsubsection{SUSY}

 The mainstream of theoretical thinking with respect to the physics of
 electroweak symmetry breaking and the physics above 100 GeV is
 dominated by those who consider that supersymmetry, SUSY, should play
 a strong role. SUSY comes in may guises but always leads to a
 proliferation of postulated particles differing in spin by one half
 unit with respect to the ``standard particles''. Thus there is a
 spin-one-half gluino, the partner of the gluon, a spin-one-half
 photino, the partner of the photon and a host of squarks and
 sleptons. The simplest assumption is that these sparticles can only
 be produced in pairs, a restriction that is usually formulated as
 conservation of a quantum number R-parity.

 Given R-parity conservation there is always a stable lightest
 supersymmetric partner(LSP) which is neutral in most theories. This
 leads to the presence of missing transverse energy as the most
 generic of SUSY indicators. As with technicolor, the use of $b$-quark
 tagging also can be a useful discriminator against background. Since
 we have not seen supersymmetric partners with the same masses as the
 ordinary particles, SUSY, if it exists must be a broken symmetry. The
 mechanism by which it is broken at very high mass scales
 distinguishes different models. Very commonly considered is the super
 gravity(SUGRA) class. In the last few years, alternatives such as
 gauge mediated models(GMSB) have been prominent. The latter are
 characterised by the gravitino($\tilde{G}$) being the LSP and
 cascades of decays such as $\chi_{1}^{0}\rightarrow \gamma\tilde{G}$
 which generate final states with one or more photons and missing
 transverse energy.

 As indicated in Table~\ref{table-susy}, the present limits for
 different sparticles range up to a couple of hundred GeV for some but
 around 100 GeV for others. The afficionados of SUSY tend to expect
 that it would appear with sparticle masses in the range below 1 TeV
 if it is to be relevant to the electroweak symmetry breaking
 problem. The recent studies\cite{susy_summary}, also summarised in
 Table~\ref{table-susy} suggest that with 2 $\rm{fb^{-1}}$ of integrated
 luminosity, these ranges can be considerably extended to cover a
 large fraction of the ``interesting'' region.


\subsection{ The Higgs Boson}

 Without looking beyond the physics of the standard model, it is
 necessary to postulate some mechanism to break the electroweak
 symmetry and to give mass to the $W$ and $Z$ bosons. The simplest
 thing to do is to assume a single complex Higgs field which in turn
 leads to a single neutral Higgs boson. As we have seen earlier the
 mass of such an object is predicted through the radiative corrections
 to the electroweak parameters and with the mass of the top quark and
 that of the W boson measured we should find the Higgs boson at the
 appropriate place.

\begin{figure}[]	
\vspace{-0.5cm}
\centerline{\epsfxsize 2.43 truein \epsfbox{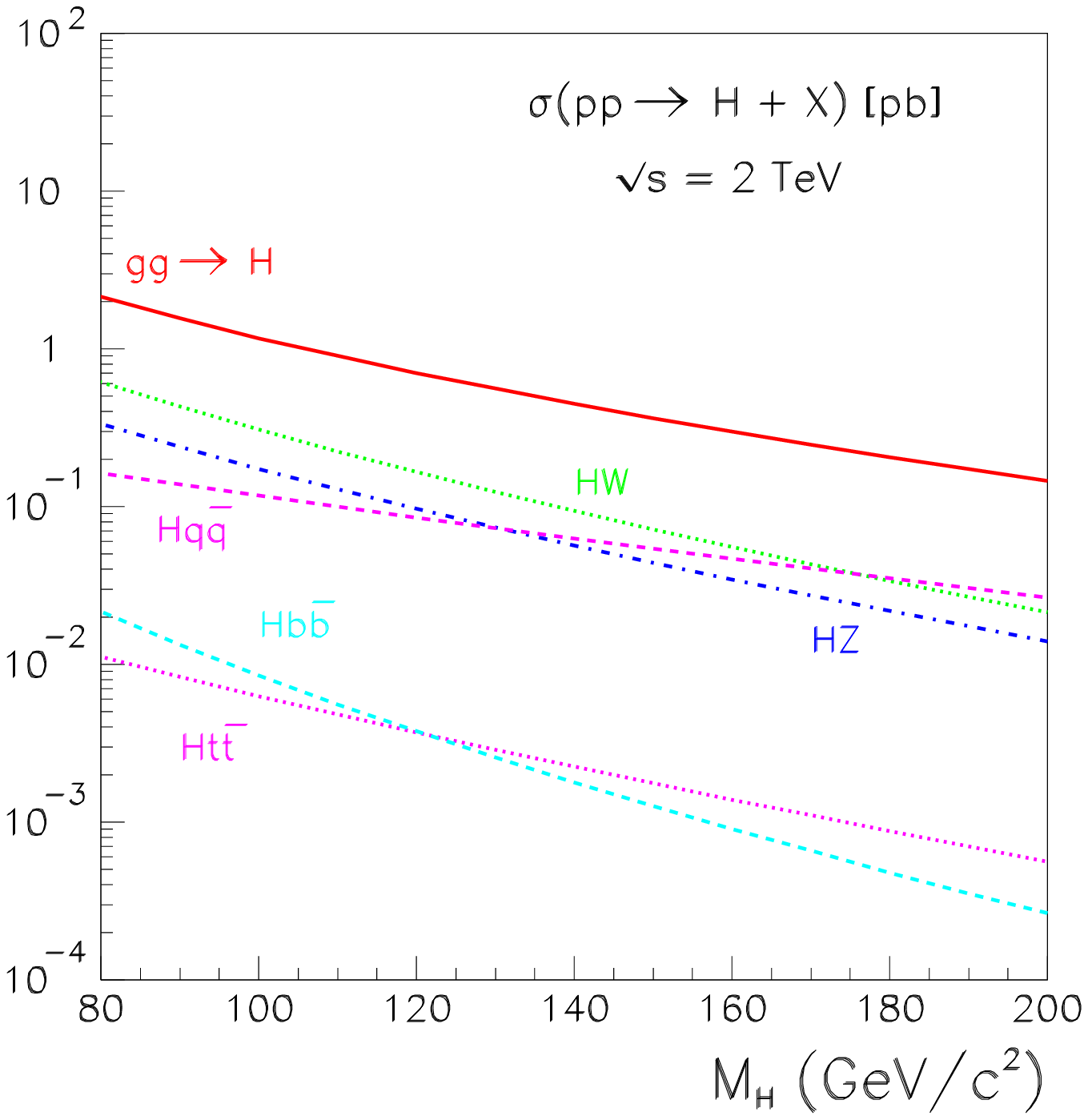} \epsfxsize 2.2 truein \epsfbox{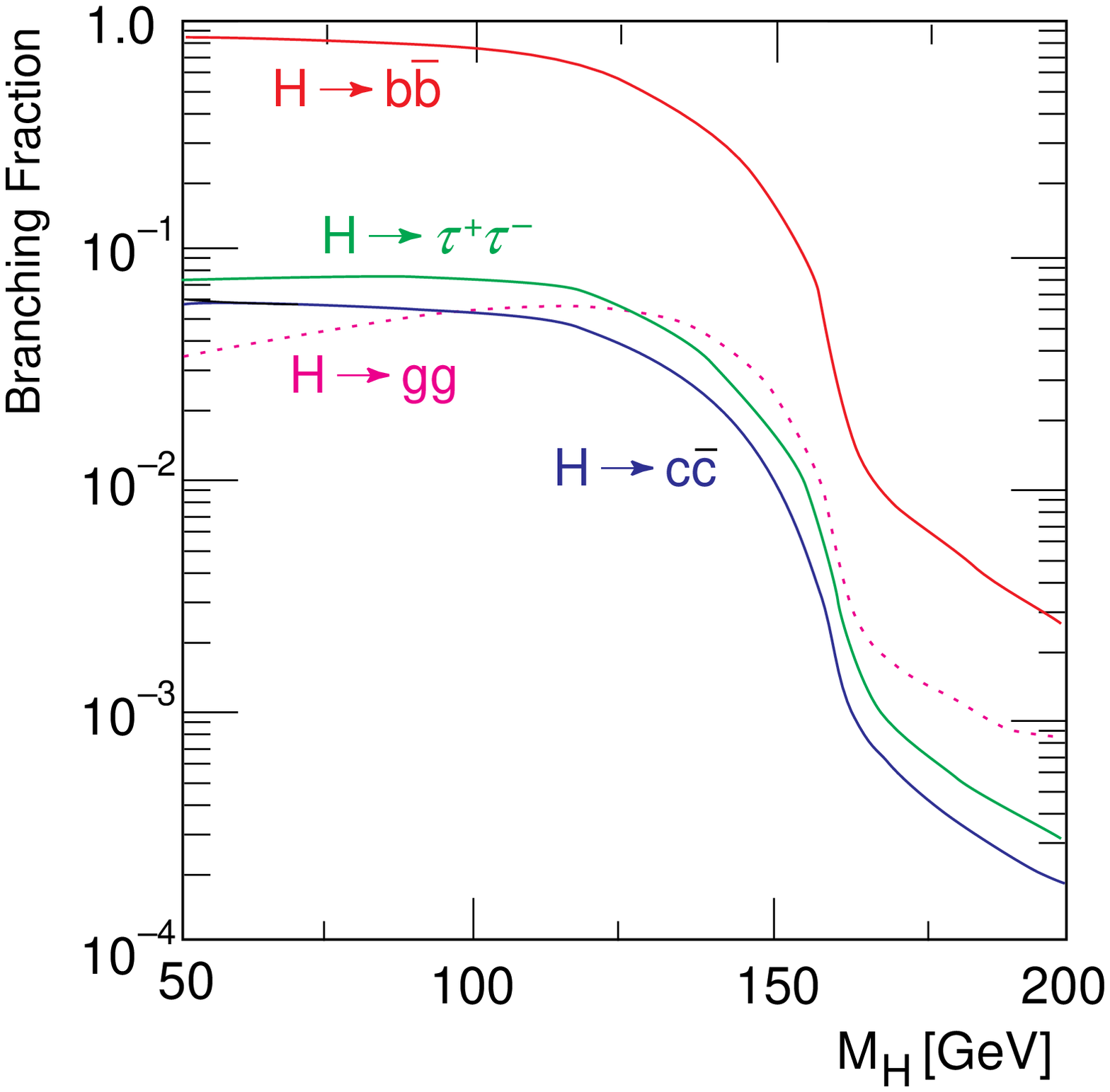}\epsfxsize 2.2 truein \epsfbox{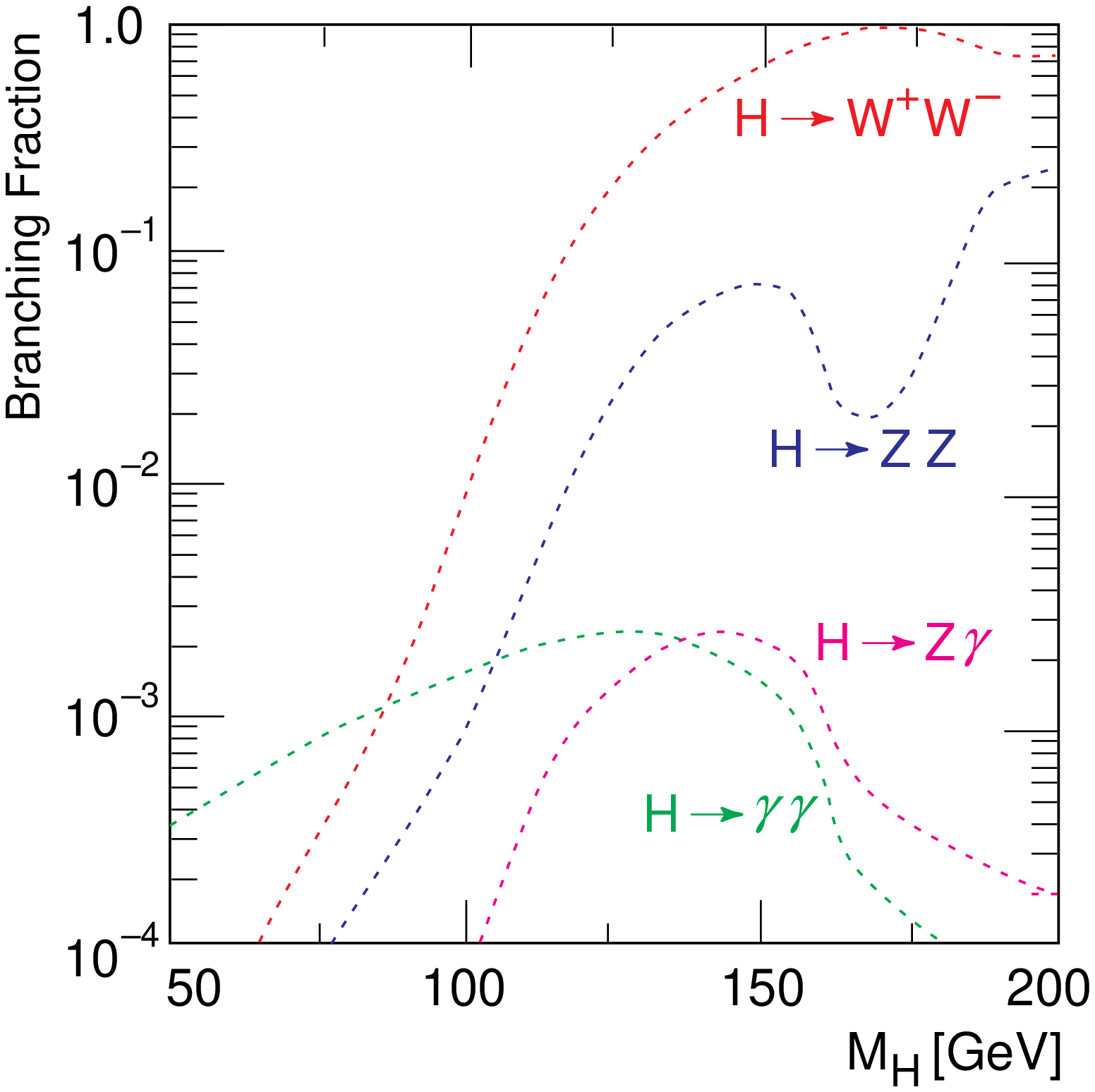} }   
\vskip 0 cm
\caption[]{
\label{fig-higgsbr}
\small Higgs-boson production cross sections and  branching fractions for fermions and bosons as a function of Higgs-boson mass.}
\end{figure}

 The search for the standard-model Higgs boson is the most widely used
 benchmark for the potential of planned collider experiments. Recent
 studies\cite{susyrun2} have put the estimates for the Tevatron
 Collider experiments, D\O\ and CDF, on a more solid footing. While
 gluon-gluon fusion has the highest Higgs production cross section,
 the associated , $WH$ and $ZH$, production channels offer distinctive
 experimental signatures through the leptonic decays of the bosons and
 have received much attention.  The $b\overline{b}$ decay of the Higgs
 also adds powerful discrimination especially at low masses, below
 about 130 GeV, where that decay dominates. A more thorough study of
 the channels involving $b$-quark tagging was conducted. Further, the
 use of the $WW$ decay modes in conjunction with the dominant
 gluon-gluon fusion has been reconsidered\cite{turcothan}.  The
 branching fractions for the latter rise strongly, see
 Fig.~\ref{fig-higgsbr}, with increasing Higgs mass and also are
 distinctive experimentally.

\begin{figure}[]	
\centerline{\epsfxsize 3.0 truein \epsfbox{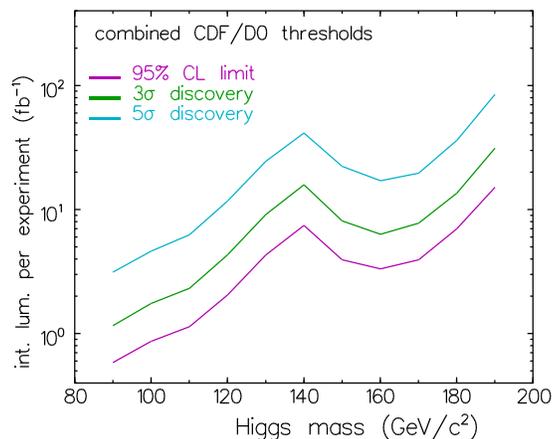}}   
\vskip 0 cm
\caption[]{
\label{fig-higgsensitivity}
\small Luminosity required as a function of Higgs mass to achieve different levels of sensitivity to the standard-model Higgs boson.From the upper curve corresponds to a $5~\sigma$ discovery, the middle a $3~\sigma$ signal and the lower a 95\% exclusion limit. These limits require two experiments, Bayesian statistics are used to combine the channels and   include the improved sensitivity which would come from multivariate analysis techniques.}
\end{figure}

The combined sensitivity for all channels and two experiments is shown
in Fig.\ref{fig-higgsensitivity}. The figure shows the required
luminosity to obtains a signal at different levels of significance,
$5~\sigma$, $3~\sigma$, and the 95\% exclusion limit, as a function of
Higgs mass. We see that with both experiments and 30 $fb^{-1}$ of
luminosity for each experiment the sensitivity extends up to Higgs
massses of 190 GeV. If the Higgs does not exist in this mass region,
with 10 $fb^{-1}$ this whole region could be excluded experimentally
at the 90\% C.L.

 Other studies concentrated on the extensions of the Higgs sector, in
 particular the SUSY-Higgs two-doublet model. There are three neutral
 and two charged Higgs states and, depending on the value of the ratio
 of the vacuum expectation values of the two doublets, very strong
 coupling of the Higgs to the $b$-quark may be expected. Again the
 r$\rm{\hat{o}}$le of the $b$-quark tagging is important and similar
 sensitivities to the SUSY Higgs are achieved as in the
 standard-model-Higgs case.

 As we discussed earlier, an integrated luminosity of 20-30 $fb^{-1}$
 before the startup of the LHC is anticipated. There is a
 considerable challenge for the experiments but even the fierce
 conditions may be mitigated by operating the machine in such a way as
 to maintain the instantaneous luminosity at or less than $5. \times
 {10^{32}} cm^{2} sec^{-1} $. 

 It is clear that maximizing the exploitation of the Tevatron Collider
 to search for the Higgs and other new high-mass physics should be one
 of the highest priorities of the U.S. high energy physics program.

\section{The Experimental Program}

 We have seen in the above discussion that the physics reach of the
 accelerator complex which we label with its newest component, The
 Fermilab Main Injector, is phenomenal and diverse. Full exploitation
 of every aspect could swallow resources in excess of what appears to
 be available.  The attempts to construct a realistic program has so
 far left the various components in states of varying certitude. While
 some pieces are well on their way to completion of construction
 others remain glints in the eyes of the proponents. In order to give
 a sense of perspective, I have chosen to give the briefest of
 summaries of the status of each of the components of the Main
 Injector physics program.

\begin{itemize}

 \item The Main Injector: the commissioning of the machine is well
 advanced; the ancillary Recycler storage ring will be completed in
 the coming months.

 \item The Tevatron Collider and the CDF and D\O\ Detectors: the
 Tevatron will operate in fixed target mode during 1999 and will
 convert to collider operations for early 2000. The upgrades of the two
 detectors CDF and D\O\ are in the middle of construction and
 completion and roll-in is expected in 2000.

 \item The NuMI Project: the project has approval from the appropriate
 authorities and a baseline for the scope, cost, funding and schedule
 has been approved by a Department of Energy review with funds for
 civil construction allocated.

 \item The mini-BooNe Experiment: this initial phase of a potentially
 longer program is an approved experiment expected to run in 2002.

 \item The Kaon-CP Violation experiments: these experiments, labelled,
 CKM, CPT and KaMI, have submitted proposals or letters of intent to
 the laboratory; research and development projects associated with
 different aspects of them have been established.

 \item The dedicated collider-B Experiment, BTeV: a letter of intent
 has been submitted and an experimental hall has been constructed; a
 research and development program is under way.

 \item A 120 GeV QCD Program: a number of groups have submitted
 proposals\cite{meson120} in response to the potential offered by
 extracted hadron beams from the the Main Injector. The primary thrust
 of such a program would be to emphasize QCD studies. Thus far there
 is no action on these proposals.

\end{itemize}

\section{Conclusions}

  The Main Injector enables a phenomenally broad and imposing array of
  physics and we, the field, must be wise in choosing which pieces to
  emphasize. Many physicists are determined to exploit the
  potential of this program. I am very excited to be among those
  physicists.

\section{Acknowledgements}

 The talk and this paper could not have been produced with out the
 help many. Included among those people are Franco Bedeschi, Ed
 Blucher, Amber Boehnlein, Greg Bock, Janet Conrad, Peter Cooper,
 Marcel Demarteau, Gene Fisk, Al Goshaw, Paul Grannis, Steve Holmes,
 Zoltan Ligeti, John Marriner, Shekhar Mishra, Meenakshi Narain, Adam
 Para, Ron Ray, Maria Roco, Ken Stanfield, Gordon Thomson, Andre
 Turcot, Harry Weerts, Bruce Winstein, Stan Wojcicki and John
 Womersley who were kind enough to provide input and/or to read a
 draft version of the paper. To them should go the credit for the
 content, to me the blame for errors.  This work was supported by the
 U.S. Department of Energy under Contract No. DE-AC02-76CHO3000.

\end{document}